\documentclass[a4paper,fleqn,usenatbib]{mnras}

\usepackage{savesym}
\usepackage{amsmath}
\savesymbol{iint}
\savesymbol{iiint}
\usepackage{txfonts}
\restoresymbol{TXF}{iint}
\restoresymbol{TXF}{iiint}

\usepackage[T1]{fontenc}
\usepackage{ae,aecompl}

\usepackage{graphicx}	

\title[Nova standard candles]{A {\it Hubble Space Telescope} survey for novae in M87. III. Are novae good standard candles 15 days after maximum brightness?}

\author[M. M. Shara et al.]{Michael M. Shara$^{1,2}$\thanks{E-mail: mshara@amnh.org}, Trisha F. Doyle$^{3}$, Ashley Pagnotta$^{1}$, James T. Garland$^{4}$, 
\newauthor {Tod~R.~Lauer$^{5}$, David~Zurek$^{1}$, Edward A. Baltz$^{6}$, Ariel Goerl$^{7}$, Attay~Kovetz$^{8}$, }
\newauthor {Tamara Machac$^{9}$, Juan P. Madrid$^{10}$, Joanna Miko{\l}ajewska$^{11}$, J. D. Neill$^{12}$, }
\newauthor {Dina~Prialnik$^{13}$, D. L. Welch$^{14}$ and Ofer~Yaron$^{15}$}\\
$^{1}$Department of Astrophysics, American Museum of Natural History, Central Park West at 79th Street, New York, NY 10024-5192 USA\\
$^{2}$Institute of Astronomy, University of Cambridge, Madingley Road, Cambridge CB3 0HA, United Kingdom\\
$^{3}$Physics and Space Sciences Department, Florida Institute of Technology, 150 W. University Boulevard, Melbourne, FL 32901-6988, USA\\
$^{4}$Professional Children's School, 132 West 60th Street, New York, NY 10023, USA\\
$^{5}$National Optical Astronomy Observatories, 950 N. Cherry Avenue, Tucson, AZ 85719-4933, USA\\
$^{6}$KIPAC, SLAC, 2575 Sand Hill Road, M/S 29, Menlo Park, CA 94025, USA\\
$^{7}$Booker T Washington Middle School, 103 West 107th Street, New York, NY 10025, USA\\
$^{8}$School of Physics and Astronomy, Sackler Faculty of Exact Sciences, Tel Aviv University, Ramat Aviv 69978, Israel\\
$^{9}$Packer Collegiate Institute, 170 Joralemon Street, Brooklyn, NY 11201, USA\\
$^{10}$CSIRO, Astronomy and Space Science, P.O. Box 76, Epping, NSW 1710, Australia\\
$^{11}$N. Copernicus Astronomical Center, Polish Academy of Sciences, Bartycka 18, PL 00--716 Warsaw, Poland\\
$^{12}$California Institute of Technology, 1200 East California Boulevard, MC 278-17, Pasadena CA 91125, USA\\
$^{13}$Department of Geophysics and Planetary Sciences, Sackler Faculty of Exact Sciences, Tel Aviv University, Ramat Aviv 69978, Israel\\
$^{14}$Department of Physics \& Astronomy, McMaster University, Hamilton, L8S 4M1, Ontario, Canada\\
$^{15}$Department of Particle Physics and Astrophysics, Weizmann Institute of Science, 76100 Rehovot, Israel\\
}

\date{Accepted XXX. Received YYY; in original form ZZZ}

\pubyear{2017}

\begin{document}
\label{firstpage}
\pagerange{\pageref{firstpage}--\pageref{lastpage}}
\maketitle

\begin{abstract}
Ten weeks of daily imaging  of the giant elliptical galaxy M87 with the {\it Hubble Space Telescope} (HST) has yielded 41 nova light curves of unprecedented quality for extragalactic cataclysmic variables. We have recently used these light curves to demonstrate that the observational scatter in the so-called Maximum-Magnitude Rate of Decline (MMRD) relation for classical novae is so large as to render the nova-MMRD useless as a standard candle. Here we demonstrate that a modified Buscombe - de Vaucouleurs hypothesis, namely that novae with decline times $t_{2}$ > 10 days converge to nearly the same absolute magnitude about two weeks after maximum light in a giant elliptical galaxy, is supported by our M87 nova data. For 13 novae with daily-sampled light curves, well determined times of maximum light in both the F606W and F814W filters, and decline times $t_{2}$ > 10 days we find that M87 novae display $M_{606W,15}$ = $-6.37 \pm 0.46 $ and $M_{814W,15}$ = $-6.11 \pm 0.43 $. If very fast novae with decline times $t_{2}$ < 10 days are excluded, the distances to novae in elliptical galaxies with stellar binary populations similar to those of M87 should be determinable with 1 sigma accuracies of $\pm 20\% $ with the above calibrations.
\end{abstract}




\section{Introduction}
Novae near maximum light are so luminous - ranging from M$_V \sim -5$ to $-10.7$ \citep{sha09} - that they can be detected with the {\it Hubble Space Telescope} (HST) out to cosmologically interesting distances, well beyond the useful ranges of RR Lyrae stars, Cepheids and planetary nebulae. After the announcement of their discovery in the Andromeda galaxy \citep{hub29}, \citet{zwi36} claimed that novae behave as standard candles, with light curves that can be interpreted to yield the distances to their host galaxies. Zwicky's initial correlation between the maximum magnitude of a nova and its rate of decline (now known as the MMRD relation), was improved upon by \citet{mcl45} and \citet{arp56}. 

If the MMRD relation did indeed calibrate novae as standard candles, then they could be used to independently measure distances to galaxies with Cepheids. More important, they could determine distances to e.g. elliptical galaxies with no Cepheids, to low luminosity dwarf galaxies \citep{con15}, and even to intergalactic `tramp novae'  \citep{sha06,tey09} which act as tracers of stars unbound to any galaxy. 

Because it is {\it not} just WD mass, but also total accreted envelope mass that control nova thermonuclear runaways, and because different accretion rates can produce a significant range of accreted envelope masses \citep{yar05,sha17}, the scatter in the MMRD was predicted to be significant \citep{pri95, yar05, hac10}. The many observed MMRD relations in the literature are summarised in \citet{dow00}. It is clear from their Figure 17 that the observational scatter in an MMRD plot is far from negligible. With the discovery of ``faint, fast novae" in M31 \citep{kas11,dar16} and in M87 \citep{sha16}, increasing the observed scatter in the MMRD to 3 magnitudes, it is now evident from observations that the above-noted predictions from theory were correct: the MMRD has no value as a distance indicator \citep{sha17}. \citet{fer03} reached the same conclusion, based on a smaller and less densely sampled group of nine novae in the Virgo cluster elliptical galaxy M49.

There is another methodology that has been proposed to use novae as distance indicators. \citet{bus55} suggested that all novae reach roughly the same absolute photographic magnitude, $M_{pg,15} \sim -5.2$ $\pm 0.1$, 14-16 days after maximum light. \citet{fer03} and \citet{dar06} presented evidence that the scatter in absolute magnitude at 15 days after maximum light for novae in M49 and in M31, respectively, is much larger than claimed by \citet{bus55}.

All of the nova magnitudes used in \citet{bus55} were, of course, photographic, rather than CCD-based. Visual and (blue) photographic magnitudes were used interchangeably in that study, using the rationale that nova colour-indices are close to zero near maximum light. Absolute magnitudes for novae were based on nebular expansion parallaxes and interstellar absorption line widths. An interstellar absorption coefficient of $A_{pg}$ = 0.8 mag/kpc was assumed to correct the measured magnitudes at maximum light. All of these assumptions were reasonable, and the best that could be done in 1955, but no quantification of the multiple sources of error was made. As noted above, the claimed error in $M_{pg,15}$, $\pm0.1$ mag, is overly optimistic \citep{fer03, dar06}.

The goals of this paper are to use our recently-reported {\it Hubble Space Telescope} (HST) survey of novae in M87, and the M87 nova light curves, to determine whether the absolute magnitudes of novae in M87 converge about 15 days after maximum light; if so, to measure those absolute magnitudes in two of the most frequently-used HST filters; and to provide realistic estimates of the errors of those absolute magnitudes.

In section 2 we summarise our observations of well-observed novae in M87, and plot the light curves and daily standard deviations of all M87 novae with well-observed maxima.  We compare those results with previous determinations in section 3, and summarise our results and conclusions in section 4.

\section{Observations and Light Curves}
We carried out {\it Hubble Space Telescope} Advanced Camera for Surveys (HST/ACS) imaging of the giant elliptical galaxy M87 in the F606W and F814W filters taken for HST Cycle-14 program 10543 (PI - E. Baltz) over the 72 day interval 24 December 2005 through 5 March 2006. Full details of the observations, data reductions, detections and characterisations of 41 variables - 32 certain and 9 likely novae - and their images, and light and colour curves, are given in \citet{sha16} (hereafter Paper I). Table 2 of that paper lists all photometric measurements (in Vega magnitudes) and errors for the 41 novae. A brief summary of the time coverage, its regularity, and the photometric errors is as follows. 

Each one-orbit epoch of observations consisted of four 360-second exposures in the F814W filter followed by one exposure of 500 seconds in the F606W filter. We note that the original motivation of this proposal was to search for microlensing events in M87, while the detection of novae was foreseen as  a potential ``bonus". That is why the bulk of exposures were taken in F814W (to detect the faintest red giants possible), and why, despite admitting H$\alpha$, the broader filter F606W was chosen rather than F555W (obviously, H$\alpha$ emission is not expected in microlensing events). Some M87 nova H$\alpha$ photons have inevitably leaked into the F606W filter so it measures both nova continuum and emission. This is a likely contributor to some of the scatter in the brightness in the F606W filter, which will vary from nova to nova, and with time for each nova.   

This survey for extragalactic novae is unique. HST enables observations with nearly perfect daily cadence and constant limiting magnitude extending over a six magnitude range throughout most of M87. Photometric errors for most novae range from 0.01 - 0.04 mag near maximum to 0.3 - 0.4 mag by the time they have faded by 2 - 3 mag from maximum.The first four epochs were spaced $5.00 \pm 0.02 $ days apart, followed by $1.00 \pm 0.11$ day spacing for the remaining 60 epochs. Thus while all novae visible on any given day were observed at precisely the same times, the maxima of individual novae were not, so the spacing between subsequent days for the novae in our sample are not all identical. As noted above, observations were taken at a cadence of  $1.00 \pm 0.11$ day. We have calculated the mean magnitudes and standard deviations of those mean nova magnitudes on day N after maximum light by using the magnitudes of each nova at the time closest to N days after its maximum. 

We considered only the 32 ``certain" novae in what follows.
1) As the survey ended, 5 of 32 novae were still visible, but all had been visible for < 15 days.
2) 8 of 32 novae were already in eruption and past maximum when the survey began.
3) 6 of 32 novae evolved so quickly, or were so close to the centre of M87, that they were too faint, 15 days after maximum light, to be detected by HST.

This left 32 - (5+8+6) = 13 novae whose maxima were seen, and which remained visible in both filters for at least 15 days.
The numbering of these novae in the figures below follows the nomenclature of Paper I. 

Novae were detected with 2-magnitude decline times $t_{2}$ ranging from 2 days to 36 days (table 3 of \citet{sha16} ). Simulations demonstrate that we detected all novae with $t_{2}$ > 2 days and which are located >  20 arcsec from the nucleus. A nova with $t_{2} \sim$1 day might have been seen only once ($\sim$1 magnitude past maximum) or twice (at $\sim$1 and $\sim$ 3 magnitudes past maximum); no such object was detected. Even faster novae ($t_{2} \sim$6 hours) are predicted to exist \citep{sha17} but at most one detection of each such object could have occurred. None was detected.

The apparent HST F606W and F814W magnitudes of the novae presented in Paper I were converted to absolute magnitudes using the distance modulus to M87 derived by \citet{bir10}, i.e. (m-M)$_0 = 31.08 \pm 0.06 $. The 16 HST F606W and 17 F814W light curves of novae in M87 with well-defined maxima are shown in Figures 1 and 2 respectively. Only those 13 novae which were visible for at least 15 days after maximum light in both filters' images were used to calculate the daily means and standard deviations of M87 nova brightnesses. The daily mean on (for example) day 7 is taken to be the mean of the magnitudes of all of the novae on the seventh day after their individual maximum. As noted above, the spacing between subsequent days for the individual novae is actually $1.00 \pm 0.11$ day. 

Except for nova 1, the novae in Figures 1 and 2 display remarkably similar light curves, and a convergence in brightness that is obvious from simple inspection. The shapes of the light curves in Figures 1 and 2 may be contrasted with those of novae in M31 \citep{dar04}, the Magellanic Clouds \citep{mro16} and in the Galaxy \citep{str10}, which display a wide range of morphologies.

\clearpage

\begin{figure}
\centering
\includegraphics[width=190mm]{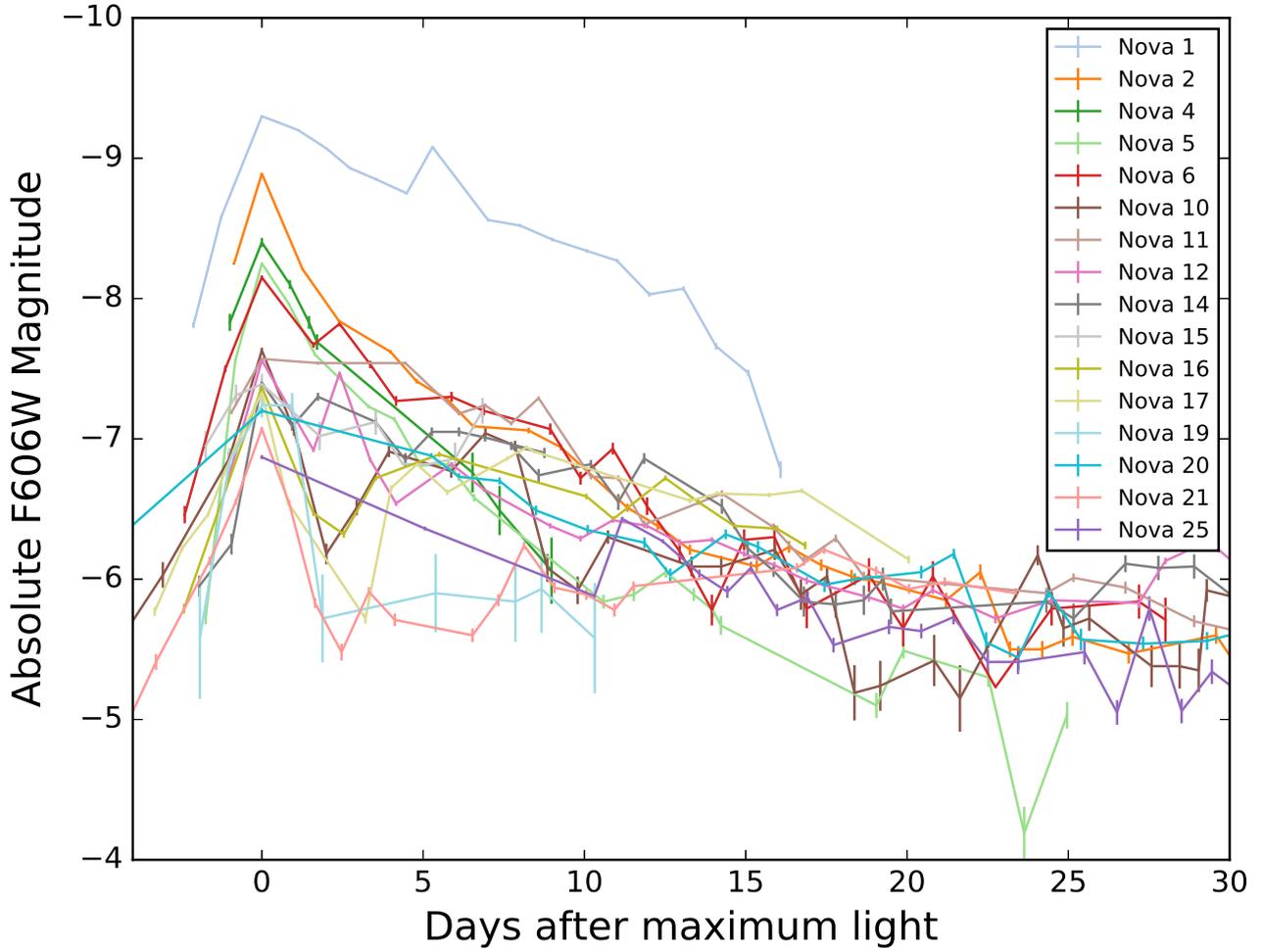}
\caption{The HST F606W light curves of 16 novae in M87 with well-observed maxima, from Paper I. Each of these nova eruptions was observed within 12 hours of maximum light. Magnitude error bars are 1 sigma.}
\end{figure}

\clearpage

\begin{figure}
\includegraphics[width=190mm]{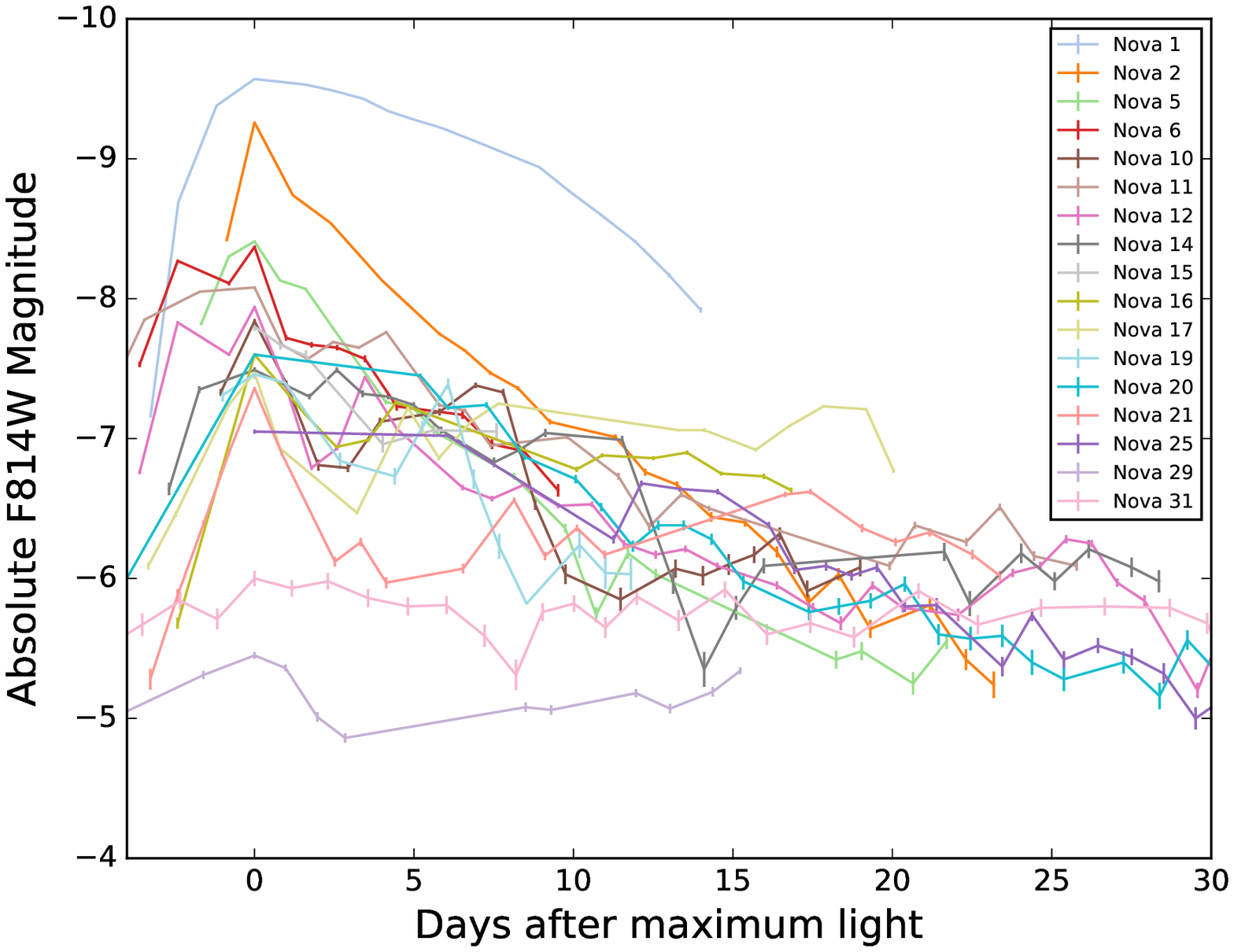}
\caption{The HST F814W light curves of 17 novae in M87 with well-observed maxima, from Paper I. Each of these eruptions was observed within 12 hours of maximum light. Magnitude error bars are 1 sigma.}
\end{figure}

\clearpage



\begin{figure}
\includegraphics[width=190mm]{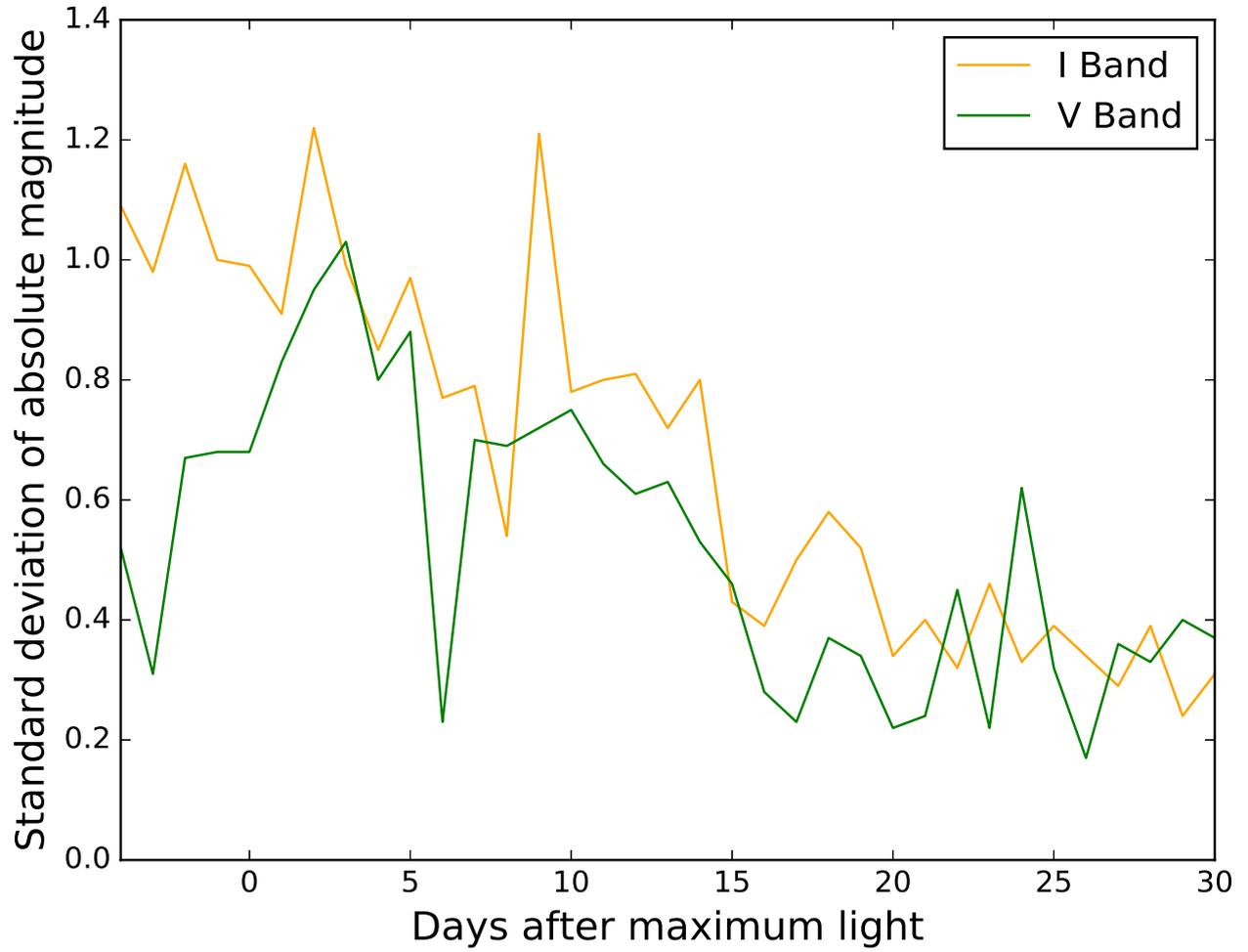}
\caption{The daily standard deviations of 13 HST F606W and F814W light curves of novae in M87 which had well-observed maxima and which were seen for at least 15 days after maximum light.}
\end{figure}

\clearpage

\begin{figure}
\includegraphics[width=190mm]{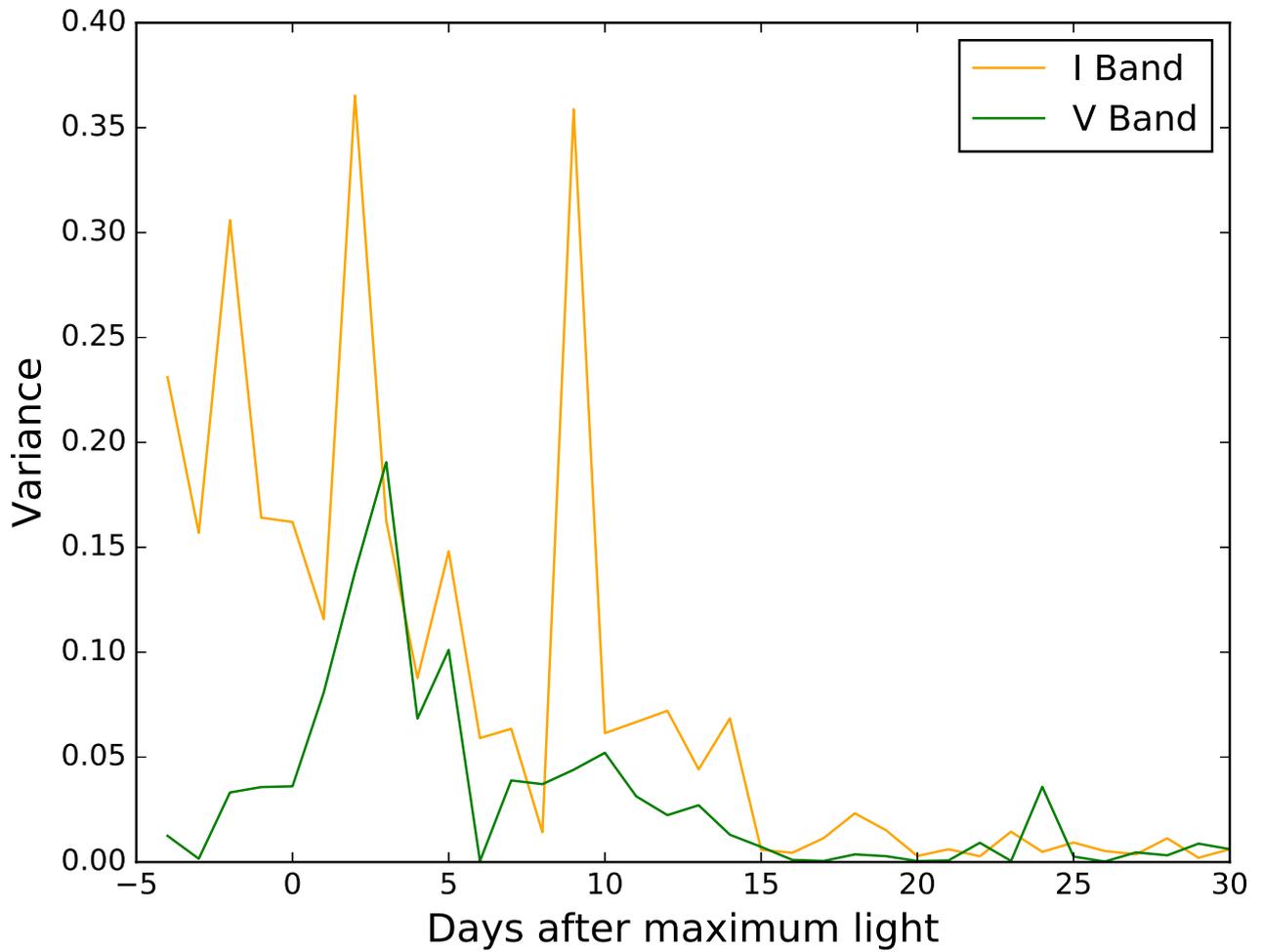}
\caption{The daily variance of the variance of 13 HST F606W and F814W light curves of novae in M87 which had well-observed maxima and which were seen for at least 15 days after maximum light.}
\end{figure}

\clearpage

\section{Using Novae as Distance Indicators}

Several authors after \citet{bus55} investigated whether novae are useful standard candles when measured 15 days after maximum brightness. 
The largest such sample of novae is from \citet{pfa76}. For 46 well-observed novae in M31, the Galaxy, and the Magellanic Clouds he determined $M_{B,15}$ = $-5.74 \pm 0.60 $. \citet{coh85} and \citet{van87} used their ground based samples of 11 and 6 Galactic novae, respectively, to determine values of $M_{V,15}$ = $-5.60 \pm 0.45 $ and $ -5.23 \pm 0.39 $. \citet{dow00} used 15 Galactic novae to determine $M_{V,15}$ = $ -6.05 \pm 0.44$.  HST observations of 5 novae in M49 allowed \citet{fer03} to measure $M_{V,15} $ = $-6.36 \pm 0.43$. They stated that their value is strongly at odds with that of \citet{van87}, barely consistent with that of \citet{coh85} and in reasonable agreement with \citet{dow00}.

\citet{dar06} found the largest scatter of all observers for their sample of M31 novae: $M_{r',15} $ = $-6.3 \pm 0.9$ and $M_{i',15} $ = $-6.3 \pm 1.0$. Their filters and effective cadence were similar to the ones used in the current work. Their result led them to write: ``We can conclude that the POINT-AGAPE CN catalogue shows no evidence of a t15 relationship, nor strong evidence of convergence at another time-scale". This is the strongest evidence and statement in the literature against using novae in spiral galaxies as standard candles when observed 15 days after maximum light. 

Figure 3 yields the epochs of minimum standard deviation of 13 M87 nova light curves observed with HST. The number of still-visible novae begins to decrease after day 16; there are, for example 10 novae left at day 21 and 7 novae left at day 29. The standard deviation is then affected by the decreasing number of novae, and a comparison with the earlier epochs is no longer meaningful.  In Figure 4 we plot the variance of the variance of the M87 nova magnitudes, which supports the \citet{bus55} empirical, photographic (pg) determination of minimum deviation at 14-16 days after maximum light, though our sample is larger, of higher cadence and based on HST CCD magnitudes. We adopt the mean magnitudes and their standard deviations of the 13 still-visible M87 novae at day 15 after maximum light as the most useful nova standard candle in an elliptical galaxy to date. 

We use the M87 distance modulus noted in section 2, and absorptions to M87 of A(F606W) = 0.07 and A(F814W) = 0.04. The absolute F606W and F814W magnitudes 15 days after maximum light are:  $M_{606W, 15}$ = $-6.37 \pm 0.46 $ and $M_{814W, 15}$ = $-6.11 \pm 0.43 $. The F606W value is in excellent agreement with the only other HST-determined value \citep{fer03}, also for an elliptical galaxy. Our M87 novae are all at the same distance, are densely time-sampled with no gaps, and are observed with invariant seeing and background that is immune to lunar phase. We thus suggest that our derived values of  $M_{F606W,15}$ and $M_{F814W,15}$ are the most reliable values currently available for using novae as standard candles in {\it elliptical galaxies}. These values of $M_{15}$, the key results of this paper, enable the determination of nova distances in elliptical galaxies with 1-sigma accuracies of $\pm 20\%$.

Finally, we note that there are theoretical studies that predict nova luminosities, and times of minimum spread of those luminosities, in good agreement with the results noted above. If one excludes the class of faint, fast novae \citep{kas11,sha17} then the absolute magnitude - rate of decline (MMRD) relation for novae (i.e. that luminous novae decline faster than faint novae) follows from the fact that nova eruptions are largely controlled by the WD mass \citep{shb81}. The scatter in MMRD is controlled by the variable envelope mass \citep{hac10}. The derivative of the MMRD relation \citet{sha81}, and the universal decline law of novae \citep{hac10} demonstrate that about 15 days after maximum light one expects all nova absolute magnitudes to decline to roughly -6.

\section{Summary and Conclusions}

We have determined that the the absolute F606W and F814W magnitudes of 13 novae in M87 observed with HST 15 days after maximum light are $M_{V,15}$ = $-6.37 \pm 0.46 $ and $M_{I,15}$ = $-6.11 \pm 0.43 $. These values are useful for determining 1-sigma nova distances in elliptical galaxies to accuracies of $\pm 20\% .$ The scatter in the magnitudes of novae in spiral galaxies, observed 15 days after maximum, is too large for them, or tramp novae, to be used as distance indicators. 

\section*{Acknowledgements}
We gratefully acknowledge the support of the STScI team responsible for ensuring timely and accurate implementation of our M87 program. Support for program \#10543 was provided by NASA through a grant from the Space Telescope Science Institute, which is operated by the Association of Universities for Research in Astronomy, Inc., under NASA contract NAS 5-26555. This research has been partly supported by the Polish NCN grant DEC-2013/10/M/ST9/00086. MMS gratefully acknowledges the support of Ethel Lipsitz and the late Hilary Lipsitz, longtime friends of the AMNH Astrophysics  department. AP, AG and TM gratefully acknowledge the support of the Richard Guilder Graduate School of the American Museum of Natural History. JTG thanks the Science Research and Mentoring Program of the American Museum of Natural History for the opportunity to participate in the research described in this paper. We thank an anonymous referee for a thorough and careful review which improved the paper substantially.




{99}

\bsp	
\label{lastpage}
\end{document}